\documentclass[twocolumn,aps,prl,amsmath,showpacs]{revtex4}

\usepackage{graphicx}

\begin{document}

\title{Coupling/decoupling between translational and rotational dynamics \\
in a supercooled molecular liquid}
\author{Song-Ho Chong$^{1}$ and Walter Kob$^{2}$}
\affiliation{
$^{1}$Institute for Molecular Science,
      Okazaki 444-8585, Japan \\
$^{2}$Laboratoire des Collo\"ides, Verres et Nanomat\'eriaux, UMR 5587 CNRS, 
      Universit\'e Montpellier 2,
      34095 Montpellier, France}
\date{\today}

\begin{abstract}

We use molecular dynamics computer simulations to investigate the coupling/decoupling
between translational and rotational dynamics in a glass-forming liquid of
dumbbells. This is done via a careful analysis of the $\alpha$-relaxation time
$\tau_{q^{*}}^{\rm C}$ of the incoherent center-of-mass density correlator at 
the structure factor peak, the $\alpha$-relaxation time $\tau_{2}$ of the reorientational 
correlator, and the translational ($D_{t}$) and rotational ($D_{r}$) diffusion constants. 
We find that the coupling between the relaxation times $\tau_{q^{*}}^{\rm C}$ 
and $\tau_{2}$ increases with decreasing temperature $T$, whereas the coupling decreases 
between the diffusivities $D_{t}$ and $D_{r}$. In addition, the $T$-dependence 
of $D_{t}$ decouples from that of $1/\tau_{2}$, which is 
consistent with previous experiments and has been interpreted as a signature
of the ``translation-rotation decoupling.'' We trace back these apparently contradicting 
observations to the dynamical heterogeneities in the system. We show that 
the decreasing coupling in the diffusivities $D_{t}$ and $D_{r}$ 
is only apparent due to the inadequacy of the concept of the rotational diffusion constant for 
describing the reorientational dynamics in the supercooled state. We also 
argue that the coupling between $\tau_{q^{*}}^{\rm C}$ and $\tau_{2}$ and the 
decoupling between $D_{t}$ and $1/\tau_{2}$, both of which strengthen upon cooling, 
can be consistently understood in terms of the growing dynamic length scale. 

\end{abstract}


\maketitle

One of the puzzling features of glass-forming systems is that upon
cooling the translational dynamics appears to decouple from the rotational
dynamics~\cite{Fujara92,Cicerone96,Andreozzi97,Ediger00,Debenedetti01}.
At high temperatures, the translational and rotational diffusion
coefficients, $D_{t}$ and $D_{r}$, respectively, are inversely
proportional to $\eta/T$, where $\eta$ is the viscosity and $T$ the
temperature, in agreement with the Stokes-Einstein (SE) and the
Stokes-Einstein-Debye (SED) relation.  Below approximately $1.2 \, T_{\rm
g}$, with $T_{\rm g}$ the glass transition temperature, the SE relation is
replaced by a fractional relation $D_{t} \propto (\eta/T)^{-\xi}$ with
$\xi < 1$, implying a more enhanced translational diffusion than 
predicted by the viscosity, whereas $D_{r}$ remains proportional to
$(\eta/T)^{-1}$ down to $T_{\rm g}$~\cite{Fujara92,Cicerone96}.  On the
other hand, recent computer simulations give evidence for a stronger
correlation between translational and rotational mobilities at lower
$T$~\cite{Mazza-all,Lombardo06}, which is apparently in contradiction
with the mentioned decoupling between $D_{t}$ and $D_{r}$.  In addition,
an enhancement of rotational diffusion relative to translation upon
cooling is reported in these simulations, which is a trend
opposite to that observed in experiments~\cite{Fujara92,Cicerone96}.
In this Letter, we report computer simulation results for gaining
more insight and unified understanding of these coupling/decoupling phenomena
in the supercooled state.

The system we consider is the binary mixture of rigid, symmetric
dumbbell molecules, denoted as $AA$ and $BB$ dumbbells, studied
in Ref.~\cite{Chong05}.  Each molecule consists of two identical
fused Lennard-Jones (LJ) particles of type $A$ or $B$ having the same
mass $m$, and their bond lengths are denoted by $l_{AA}$ and
$l_{BB}$. The interaction between two molecules is given by the sum of the
LJ interactions $V_{\alpha \beta}(r)$ between the four constituent sites, 
with the LJ parameters $\epsilon_{\alpha \beta}$ and $\sigma_{\alpha \beta}$
for $\alpha, \beta \in \{A,B\}$ taken from Ref.~\cite{Kob94}, which are
slightly modified so that $V_{\alpha \beta}(r)$ and $V_{\alpha \beta}^{\prime}(r)$ 
are zero at the cutoff $r_{\rm cut} = 2.5 \sigma_{\alpha \beta}$ 
(see Ref.~\cite{Chong05} for details). Bond lengths are specified
by a parameter $\zeta \equiv l_{AA} / \sigma_{AA} = l_{BB} /
\sigma_{BB}$, and a sufficiently large value $\zeta = 0.8$ is chosen
so that anomalous reorientational dynamics caused by the so-called type-A
transition is absent~\cite{Chong05,Chong-HDS}.  The number of $AA$ and
$BB$ dumbbells is $N_{AA} = 800$ and $N_{BB} = 200$.  In the following
all quantities are expressed in reduced units with the unit of length
$\sigma_{AA}$, the unit of energy $\epsilon_{AA}$ (setting $k_{B} =
1$), and the unit of time $(m \sigma_{AA}^{2} / \epsilon_{AA})^{1/2}$.
Standard molecular dynamics simulations have been performed as in
Ref.~\cite{Kob94} with the cubic box of length $L = 10.564$ for $2.0
\le T \le 10$.  The longest runs were $2 \times 10^{9}$ time steps, and
we performed 16 independent runs to improve the statistics.  Such long
simulations were necessary to reach below the critical temperature
$T_{\rm c}$ of the mode-coupling theory~\cite{Goetze92}, which
we estimate as $T_{\rm c} \approx 2.10$ based on an analysis 
similar to the one done in 
Ref.~\cite{Kob94}.  In experiments $T_{\rm c}$ is found to be 
$\approx 1.2 \, T_{\rm g}$~\cite{Roessler90}, i.e., close to the temperature at
which the aforementioned decoupling sets in~\cite{Fujara92}.  We also
introduce the onset temperature, found to be $T_{\rm onset} \approx 4.0$,
below which correlators exhibit the two-step relaxation, a characteristic
feature of the glassy dynamics in which molecules are temporarily
caged by their neighbors.  Hereafter, all quantities 
used in our discussion refer to those for $AA$ dumbbells, and the
subscript $AA$ shall be dropped for notational simplicity.

In the present study we will characterize the translational dynamics 
by the incoherent center-of-mass density correlator $F_{q}^{\rm C}(t) =
(1/N) \sum_{j} \langle e^{i {\vec q} \cdot [{\vec r}_{j}^{\, {\rm C}}(t)
- {\vec r}_{j}^{\, {\rm C}}(0)]} \rangle$, and the rotational dynamics by 
$C_{\ell}(t) = (1/N) \sum_{j} \langle P_{\ell}[ {\vec e}_{j}(t) \cdot {\vec e}_{j}(0)
] \rangle$.  Here ${\vec r}_{j}^{\, {\rm C}}(t)$ and ${\vec e}_{j}(t)$
denote the center-of-mass position and the orientation of the $j$th
molecule at time $t$, respectively, and $P_{\ell}$ is the Legendre
polynomial of order $\ell$.  
The $\alpha$-relaxation times $\tau_{q}^{\rm C}$ and $\tau_{\ell}$
shall be defined via $F_{q}^{\rm C}(\tau_{q}^{\rm C}) = 0.1$
and $C_{\ell}(\tau_{\ell}) = 0.1$.

Previous simulation studies have used various ways to classify particle
mobility~\cite{Mazza-all,Lombardo06,Qian99,Berthier04,Appignanesi06}.  In this Letter,
translational and rotational mobilities of individual molecule shall be classified in
terms of the {\em first passage times}, $\tau_{q,j}^{\rm C}$ and
$\tau_{\ell,j}$, at which individual-molecule quantities $e^{i {\vec q}
\cdot [{\vec r}_{j}^{\, {\rm C}}(t) - {\vec r}_{j}^{\, {\rm C}}(0)]}$
(averaged over ${\vec q}$ having the same modulus $q$)
and $P_{\ell}[ {\vec e}_{j}(t) \cdot {\vec e}_{j}(0) ]$ become zero for
the first time.  We confirmed that $\tau_{q}^{\rm C} \propto (1/N)
\sum_{j} \tau_{q,j}^{\rm C}$ and $\tau_{\ell} \propto (1/N) \sum_{j}
\tau_{\ell,j}$ hold, and hence, $\tau_{q,j}^{\rm C}$ and $\tau_{\ell,j}$
can also be regarded as individual molecule's relaxation times. 
In the following, we will mainly refer to the structure factor peak position
$q^{*} = 8.0$ and $\ell = 2$ because of their experimental 
significance~\cite{Fujara92,Cicerone96,Mezei87-Yamamoto98}.

\begin{figure}[tb]
\includegraphics[width=0.8\linewidth]{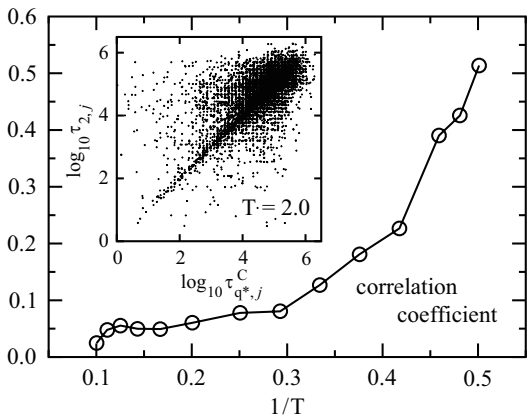}
\caption{ 
Correlation coefficient between $\tau_{q^{*}\!,j}^{\rm C}$
and $\tau_{2,j}$ as a function of $1/T$.  The Inset is a scatter
plot of $\tau_{2,j}$ versus $\tau_{q^{*}\!,j}^{\rm C}$ for $T = 2.0$
on double logarithmic scales.
}
\label{fig:correlation-coeff}
\end{figure}

The Inset of Fig.~\ref{fig:correlation-coeff} shows a  scatter plot of
$\tau_{2,j}$ versus $\tau_{q^{*}\!,j}^{\rm C}$ for $T=2.0$. We recognize
that there is a strong correlation between these two quantities, in
particular for the molecules that move/rotate quickly. From such a scatter
plot we can calculate the correlation coefficient and its $T$-dependence
is shown in the main panel of Fig.~\ref{fig:correlation-coeff}.  We see
that the correlation between translation and rotation increases 
quickly with decreasing $T$, showing the strong coupling between these
two types of motions at low temperatures, in agreement with the findings
in recent simulation studies~\cite{Mazza-all,Lombardo06}.

We also mention that the molecules that have a high translational
mobility are clustered in space, in agreement with the findings for
other glass formers~\cite{Lombardo06,Berthier04,Appignanesi06}, and the same is true for
those rotating quickly. This result is in accord with the
correlation shown in Fig.~\ref{fig:correlation-coeff}.

We next consider the translational ($D_{t}$) and
rotational ($D_{r}$) diffusion coefficients.  $D_{t}$ is determined via
$D_{t} = \lim_{t \to \infty} \Delta r_{\rm C}^{2}(t) / 6t$ of
the center-of-mass mean-squared displacement $\Delta r_{\rm C}^{2}(t) 
= (1/N) \sum_{j} \langle [ {\vec r}_{j}^{\, {\rm C}}(t) - {\vec r}_{j}^{\,
{\rm C}}(0) ]^{2} \rangle$. $D_{r}$ is calculated using the Einstein
formulation~\cite{Kaemmerer97}, i.e., via $D_{r} = \lim_{t \to \infty}
\Delta \phi^{2}(t) / 4t$ of the mean-squared angular displacement
\begin{equation}
\Delta \phi^{2}(t) = (1/N) \sum_{j} \sum_{\alpha = X,Y} 
\langle \Delta \phi_{j, \alpha}(t)^{2} \rangle.
\label{eq:delta-phi-squared}
\end{equation}
Here $\Delta \phi_{j, \alpha}(t) = \int_{0}^{t} dt' \, \omega_{j,
\alpha}(t')$ with $\omega_{j,\alpha}(t)$ denoting the $\alpha$-component
of the angular velocity of the $j$th molecule.  Note that for linear
molecules, the summation over $\alpha$ is for the two directions
$X$ and $Y$ perpendicular to the $Z$ axis chosen along ${\vec
e}_{j}(0)$~\cite{Lynden-Bell-all}.

\begin{figure}[tb]
\includegraphics[width=0.9\linewidth]{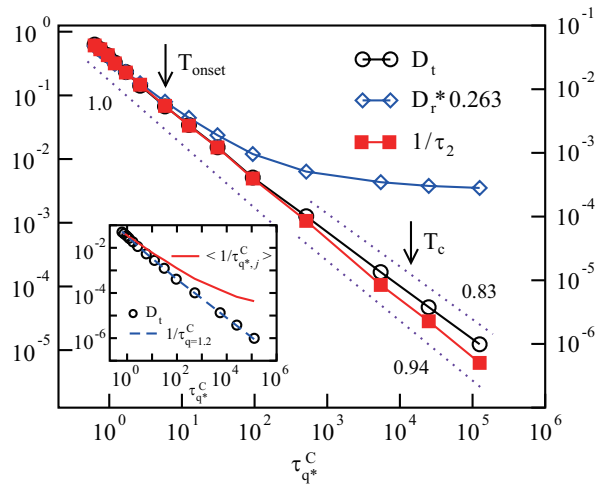}
\caption{
(Color online) Log-log plot of the translational diffusion coefficient
$D_{t}$ (circles, right scale), the rotational diffusion coefficient
$D_{r}$ (diamonds, right scale, shifted so that $D_{t}$ and $D_{r}$
agree at high $T$), and the inverse of the rotational relaxation time
$1/\tau_{2}$ (filled squares, left scale) versus the $\alpha$-relaxation
time $\tau_{q^{*}}^{\rm C}$.  The dotted straight lines refer
to $(\tau_{q^{*}}^{\rm C})^{-x}$ with the exponents $x$ cited in
the figure.  The arrows indicate the locations of $T_{\rm onset}$ and
$T_{\rm c}$.
Inset: $D_{t}$ (circles) versus $\tau_{q^{*}}^{\rm C}$ 
from the main panel is compared with 
$\langle 1/\tau_{q^{*}\!,j}^{\rm C} \rangle$ (solid line)
and $1/\tau_{q}^{\rm C}$ for $q = 1.2$ (dashed line), which are
vertically shifted so that they agree with $D_{t}$ at high $T$.}
\label{fig:Fujara}
\end{figure}

Figure~\ref{fig:Fujara} shows the log-log plot of $D_{t}$ (circles)
and $D_{r}$ (diamonds) versus $\tau_{q^{*}}^{\rm C}$. Thus, this graph
mimics Fig.~5 of Ref.~\cite{Fujara92} where $D_{t}$ and $D_{r}$ for
orthoterphenyl are traced as a function of $\eta/T$. We have used
$\tau_{q^{*}}^{\rm C}$ as a substitute for $\eta/T$ since accurate
calculation of $\eta$ is computationally challenging. However, this
approximation should be quite accurate since the $\alpha$-relaxation
time at the structure factor peak position is known to track
$\eta/T$~\cite{Mezei87-Yamamoto98}.

From Fig.~\ref{fig:Fujara} we see that $D_{t} \propto (\tau_{q^{*}}^{\rm
C})^{-1}$ holds at high $T$, whereas it is replaced by a fractional
relation $D_{t} \propto (\tau_{q^{*}}^{\rm C})^{-0.83}$ for $T \lesssim
T_{\rm c}$.  This anomalous behavior (the breakdown
of the SE relation) and its occurrence near $T_{\rm c} \approx 1.2 \,
T_{\rm g}$ are in agreement with experiments~\cite{Fujara92}.  For $D_{r}$
we recognize that $D_{r} \propto (\tau_{q^{*}}^{\rm C})^{-1}$, and hence
$D_{r} \propto D_{t}$, hold at high $T$, but that this quantity strongly
decouples from $D_{t}$ for $T < T_{\rm onset}$. That the $T$-dependence of
$D_{r}$ is weaker than the one of $D_{t}$ is in contradiction to the trend
observed in experiments~\cite{Fujara92}, but is in agreement
with other recent simulation studies~\cite{Mazza-all,Lombardo06}.

Let us notice that the experiment in Ref.~\cite{Fujara92} does not
directly probe $D_{r}$, but determines $D_{r}$ from the measurement of
$\tau_{2}$ via the relation $D_{r} \propto 1/\tau_{2}$ predicted by the
Debye model. Thus, what is reported as $D_{r}$ in Ref.~\cite{Fujara92}
corresponds in fact to $1/\tau_{2}$. In Fig.~\ref{fig:Fujara} we have
also included our simulation results for $1/\tau_{2}$ and we see that
$\tau_{2}$ correlates well with $\tau_{q^{*}}^{\rm C}$, in accord with our
previous discussion for the relaxation times, although a slight deviation
from the strict proportionality is discernible for $T \lesssim T_{\rm c}$.  
Thus, by identifying $\tau_{q^{*}}^{\rm C}$ with
$\eta/T$, our simulation result for $1/\tau_{2}$ is in accord with
the experiment~\cite{Fujara92} in which one finds $1/\tau_{2} \propto
(\eta/T)^{-1}$ down to $T_{\rm g}$. Therefore, the decoupling of $D_{r}$
from $1/\tau_{2}$ shown in Fig.~\ref{fig:Fujara} only indicates that
the Debye model breaks down for $T < T_{\rm onset}$, 
and hence, $1/\tau_{2}$ should not be considered as being proportional to $D_{r}$ 
in the supercooled state. 

\begin{figure}[tb]
\includegraphics[width=0.8\linewidth]{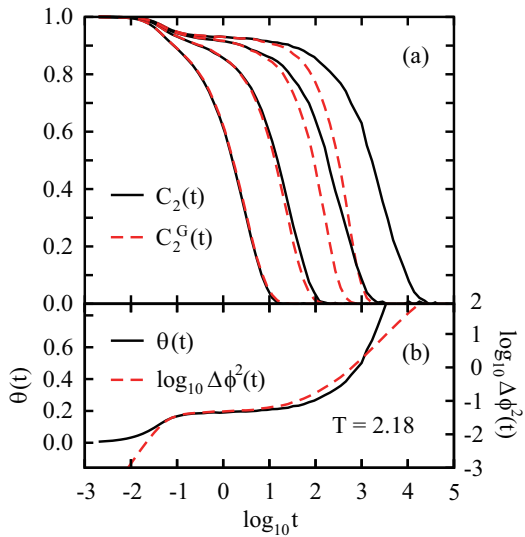}
\caption{
(Color online) (a) The reorientational correlator $C_{2}(t)$ (solid lines)
and the one calculated within the Gaussian approximation $C_{2}^{\rm
G}(t)$ (dashed lines) versus $\log_{10} t$  for $T = 5.0$, 3.0, 2.39,
and 2.18 (from left to right).  (b) The average orientation $\theta(t)$
(left scale) and the logarithm of the mean-squared angular displacement
$\log_{10} \Delta \phi^{2}(t)$ (right scale) for $T = 2.18$.
}
\label{fig:C2s-vs-Gaussian}
\end{figure}

To understand why $D_{t}$ and $D_{r}$ behave so differently, we analyze
the breakdown of the Debye model in more detail.  Within the Gaussian
approximation one has $C_{\ell}^{\rm G}(t) = \exp[ - \ell (\ell+1)
\Delta \phi^{2}(t) / 4 ]$~\cite{Lynden-Bell-all}.  The long-time
behavior is given by $C_{\ell}^{\rm G}(t) = \exp[ - \ell (\ell+1)
D_{r} t]$, which agrees with the prediction from the Debye model.
From Fig.~\ref{fig:Fujara} one therefore expects that the Gaussian
approximation breaks down for $T < T_{\rm onset} \approx 4.0$, and this
is indeed the case as shown in Fig.~\ref{fig:C2s-vs-Gaussian}a.

Figure~\ref{fig:C2s-vs-Gaussian}a also implies that for $T < T_{\rm
onset}$ there exists a time window during which $\Delta \phi^{2}(t)$
increases significantly although the orientation of a molecule
hardly changes.  The existence of such a window is demonstrated in
Fig.~\ref{fig:C2s-vs-Gaussian}b for a representative temperature,
from which one recognizes that $\Delta \phi^{2}(t)$ starts to increase
noticeably at $\log_{10} t \approx 1$ whereas the average orientation $\theta(t) =
(1/N) \sum_{j} \langle \cos^{-1}[ {\vec e}_{j}(t) \cdot {\vec e}_{j}(0)
] \rangle$ remains nearly constant up to $\log_{10} t \approx 2$.

The appearance of the plateau in $C_{2}(t)$ and $\theta(t)$ for $T <
T_{\rm onset}$ reflects the librational motion of a molecule trapped
inside the cage, with the orientation ${\vec e}_{j}(t)$ of the molecule
remaining close to ${\vec e}_{j}(0)$.  (For the translational degrees
of freedom this corresponds to rattling of the center-of-mass position.)
If the average orientation during the librational motion coincides with
${\vec e}_{j}(0)$, this gives rise to a diffusive movement of $\Delta
\phi_{j,Z}(t)$~\cite{Kaemmerer97}, which however does not contribute to
$\Delta \phi^{2}(t)$, see Eq.~(\ref{eq:delta-phi-squared}).  If, on the
other hand, the average orientation is tilted from ${\vec e}_{j}(0)$,
one can show that $\Delta \phi_{j,X}(t)$ and $\Delta \phi_{j,Y}(t)$ also
exhibit a diffusive movement.  Thus, a diffusive increase in $\Delta
\phi^{2}(t)$ occurs even if the molecule has not yet reoriented and is
still performing the librational motion inside the cage, and this effect
explains the difference between $\log_{10} \Delta \phi^{2}(t)$ and $\theta(t)$ at
intermediate times as seen in Fig.~\ref{fig:C2s-vs-Gaussian}b.  $D_{r}$
from the Einstein formulation is thus more affected by the librational
motion than the real ``reorientational'' dynamics, leading to a spurious
``decoupling'' between $D_{t}$ and $D_{r}$.
Since there is no other proper way to define $D_{r}$, we conclude that
the concept of the rotational diffusion constant is inadequate
for describing the reorientational dynamics in the supercooled state.

Now let us turn our attention to the decoupling between $1/\tau_{q^{*}}^{\rm C}$ and
$D_{t}$, and between $1/\tau_{2}$ and $D_{t}$, 
which occurs near and below $T_{\rm c}$ (see Fig.~\ref{fig:Fujara}).
The latter is known as the ``translation-rotation decoupling.''
It was conjectured that such a decoupling arises because $1/\tau_{2}$ and $D_{t}$
reflect different moments of the distribution of relaxation times, 
i.e., $1/\tau_{2} \propto 1/\langle \tau_{2,j} \rangle$ while
$D_{t} \propto \langle 1/\tau_{q^{*}\!,j}^{\rm C} \rangle$~\cite{Ediger00}.
The essential ingredient of this conjecture is the assumption that local translational mobility is 
proportional to local rotational mobility~\cite{Ediger00}, which is justified from 
our simulation (see Fig.~\ref{fig:correlation-coeff}).
However, while $\tau_{2} \propto \langle \tau_{2,j} \rangle$ has been
confirmed from our simulation, the $T$-dependence of
$\langle 1/\tau_{q^{*}\!,j}^{\rm C} \rangle$ is not in accord with that of
$D_{t}$ as demonstrated in the Inset of Fig.~\ref{fig:Fujara}.
On the other hand, as also shown there, the inverse of the $\alpha$-relaxation
time $1/\tau_{q}^{\rm C}$ for $q = 1.2$ is found to track $D_{t}$
in the simulated $T$ range. 

\begin{figure}[tb]
\includegraphics[width=0.8\linewidth]{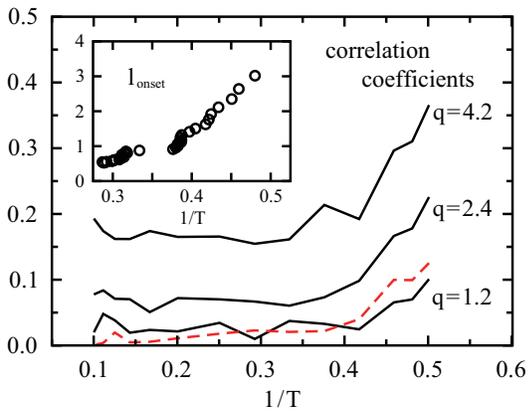}
\caption{
(Color online) Correlation coefficients between $\tau_{q^{*}\!,j}^{\rm C}$
and $\tau_{q,j}^{\rm C}$'s (solid lines) for $q = 4.2$, 2.4, and 1.2 versus $1/T$. 
The dashed line denotes the correlation coefficient between
$\tau_{2,j}$ and $\tau_{q=1.2,j}^{\rm C}$.
The Inset exhibits the length scale $l_{\rm onset}$ introduced in the text
as a function of $1/T$.}
\label{fig:length}
\end{figure}

That $D_{t} \propto (\tau_{q=1.2}^{\rm C})^{-1}$ while
$D_{t} \propto (\tau_{q^{*}}^{\rm C})^{-0.83}$ for $T \lesssim T_{\rm c}$
can be understood in terms of the growing onset length scale $l_{\rm onset}$ of Fickian
diffusion~\cite{Berthier05,Chong08}. 
Here, $l_{\rm onset}$ is defined so that the relaxation time 
$\tau_{q}^{\rm C}$ for $2\pi/q < l_{\rm onset}$ decouples from 
the diffusivity $D_{t}$, and can be estimated from the ratio $R_{q}^{\rm C}$ 
of the product $q^{2} D_{t} \tau_{q}^{\rm C}$ 
to the one at some reference temperature $T_{\rm ref}$ 
with a procedure detailed in Ref.~\cite{Chong08}. 
(In the present work we have chosen $T_{\rm ref} = 5.0$ and a criterion
$R_{q}^{\rm C} = 1.2$ for determining $l_{\rm onset}$.)
The resulting $l_{\rm onset}$ as a function of $1/T$ is presented in the
Inset of Fig.~\ref{fig:length}, from which one finds $l_{\rm onset} \approx 3$
for $1/T \approx 1/T_{\rm c} = 0.48$.
Thus, $D_{t} \propto (\tau_{q=1.2}^{\rm C})^{-1}$ for $T \lesssim T_{\rm c}$
since the associated length scale $2\pi/1.2 \approx 5.2$ exceeds $l_{\rm onset}$,
whereas $\tau_{q^{*}}^{\rm C}$ decouples from $D_{t}$ since 
$2\pi/q^{*} \approx 0.8$ is smaller than $l_{\rm onset}$.

To what extent the dynamics occurring on different length
scales are correlated is examined in the main panel of Fig.~\ref{fig:length}, 
where correlation coefficients between $\tau_{q^{*}\!,j}^{\rm C}$ and
$\tau_{q,j}^{\rm C}$'s for $q = 4.2$, 2.4, and 1.2 are plotted with
solid lines. 
The mentioned decoupling between $\tau_{q^{*}}^{\rm C}$ and $\tau_{q=1.2}^{\rm C}$
is reflected in the small value ($\lesssim 0.1$) of the corresponding
correlation coefficient. 
On the other hand, one infers from Fig.~\ref{fig:length} that
a significant correlation, characterized, e.g., by the correlation
coefficient exceeding 0.2, 
develops between $\tau_{q^{*}\!,j}^{\rm C}$ and 
$\tau_{q=4.2,j}^{\rm C}$ for $1/T \gtrsim 0.4$,
where $l_{\rm onset}$ becomes larger than the associated length 
scale $2\pi/4.2 \approx 1.5$.
Similarly, one finds that the correlation coefficient between
$\tau_{q^{*}\!,j}^{\rm C}$ and $\tau_{q=2.4,j}^{\rm C}$ 
exceeds 0.2 only for $1/T \gtrsim 0.5$, where $l_{\rm onset}$
is larger than $2 \pi/2.4 \approx 2.6$.
Thus, though originally introduced as the onset length scale of Fickian
diffusion, $l_{\rm onset}$ also provides a coherent length 
in the sense that the dynamics occurring on length scales smaller
than $l_{\rm onset}$ develop significant correlations.

Let us estimate the length scale associated with $\tau_{2}$ utilizing such a 
``coherent length'' $l_{\rm onset}$.
One finds from Fig.~\ref{fig:correlation-coeff} that the correlation coefficient
between $\tau_{q^{*}\!,j}^{\rm C}$ and $\tau_{2,j}$ exceeds 0.2
for $1/T \gtrsim 0.4$.
According to the mentioned interpretation of $l_{\rm onset}$, the dynamics
responsible for $\tau_{q^{*}}^{\rm C}$ and $\tau_{2}$ become correlated
if both of their characteristic length scales become smaller than $l_{\rm onset}$. 
Thus, the length scale associated with $\tau_{2}$ can be estimated as $\approx 1.5$
which is determined by $l_{\rm onset}$ at $1/T \approx 0.4$ (see Fig.~\ref{fig:length}).
This explains why the correlation coefficient between $\tau_{2,j}$ and
$\tau_{q=1.2,j}^{\rm C}$ is small ($\lesssim 0.1$) as shown with the
dashed line in Fig.~\ref{fig:length}, i.e., why $1/\tau_{2}$ decouples from 
$(\tau_{q=1,2}^{\rm C})^{-1} \propto D_{t}$.

Thus, the translation-rotation decoupling between $D_{t}$ and $\tau_{2}$
should more properly be understood as the decoupling of the dynamics
occurring on different length scales, which arises due to the 
growing dynamic length scale. 
This notion also explains the decoupling between
$\tau_{q^{*}}^{\rm C}$ and $D_{t}$ (the breakdown of the SE relation),
as well as the translation-rotation ``coupling''
between $\tau_{q^{*}}^{\rm C}$ and $\tau_{2}$ occurring on comparable 
length scales. 

We thank W. G\"otze for discussions. 
This work has been supported by 
Grant-in-Aids for scientific 
research from the 
Ministry of Education, Culture, Sports, Science and 
Technology of Japan (No.~20740245).

\end{document}